\documentclass[preprint,prb,showpacs]{revtex4}

\usepackage{graphicx}
\usepackage{dcolumn}
\usepackage{bm}
\usepackage{natbib}
\usepackage[english]{babel}

\newcommand{\comment}[1]{}

\begin{document}

\title{\boldmath Electron-boson spectral density functions of cuprates obtained from optical spectra via machine learning \unboldmath}

\author{Hwiwoo Park$^1$} \author{Jun H. Park$^2$} \author{Jungseek Hwang$^{1}$}\email{jungseek@skku.edu}
\affiliation{$^1$Department of Physics, Sungkyunkwan University, Suwon, Gyeonggi-do 16419, Republic of Korea\\ $^2$School of Mechanical Engineering, Sungkyunkwan University, Suwon, Gyeonggi-do 16419, Republic of Korea}

\date{\today}

\begin{abstract}

The electron-boson spectral density (EBSD) function can be obtained from measured optical scattering rate by solving a generalized Allen formula, which relates the two quantities with an integral equation and is an inversion problem. Thus far, numerical approaches, such as the maximum entropy method (MEM) and the least squares fitting method, have been applied for solving the generalized Allen formula. Here, we developed a new method to obtain the EBSD functions from the optical scattering rate using a machine learning approach (MLA). We found that the MLA is more robust against random noise compared with the MEM. We applied the new developed MLA to experimentally measured optical scattering rates and obtained reliable EBSD functions in terms of their shapes including the amplitudes. We expect that the MLA can be a useful and rapid method for solving other inversion problems, which may contain random noise.

\end{abstract}

\maketitle

\section{Introduction}
Electron-phonon spectral density (EPSD) function [$\alpha^2F(\omega)$] is a very important physical quantity for understanding superconductivity in the conventional Bardeen-Cooper-Schrieffer (BCS) superconductors\cite{carbotte:1990}. Here, $\alpha$ represents the coupling constant between an itinerant electron and the force-mediating phonon and $F(\omega)$ represents the phonon spectrum. Therefore, this quantity can carry information on the pairing interaction for forming the Cooper pairs. The EPSD function of Pb is obtained by using experimental techniques including optical spectroscopy and a theoretical method\cite{mcmillan:1965,farnworth:1976,tomlinson:1976}. The EPSD functions of Pb obtained by using different experimental spectroscopic techniques and theoretical method agree with each other and provide the correct superconducting transition temperature ($T_c$) of Pb, indicating that the phonons play the role of pairing glue for forming the Cooper pairs\cite{farnworth:1974}.

To understand superconductivity in unconventional copper oxide superconductors (cuprates), many researchers have made considerable efforts to extract the corresponding pairing electron-boson spectral density (EBSD) function from experimentally measured spectra using various experimental spectroscopic techniques including optical spectroscopy\cite{carbotte:2011}. The resulting EBSD functions showed generic temperature- and doping-dependent properties; at high temperatures above $T_c$, the EBSD functions show a broad spectrum of bosonic excitations, which extended over a wide spectral range above 400 meV, and evolved into a peak in 30-60 meV region and a featureless high-frequency background\cite{carbotte:2011} at low temperatures near or below $T_c$. This behavior is consistent with that of the spin fluctuation spectrum measured by the inelastic neutron scattering experiment. In particular, the EBSD functions obtained from measured optical spectra are crucial, because optical spectroscopy can be used to investigate almost all cuprate systems\cite{dordevic:2005,hwang:2006,hwang:2007,hwang:2008c,heumen:2009,heumen:2009a,yang:2009,hwang:2011,hwang:2013}. Optical spectroscopy can also be a bridge experimental technique between spectroscopic experimental techniques such as angle-resolved photoemission spectroscopy (ARPES), scanning tunneling microscopy (STM), and inelastic neutron scattering (INS) because it is not very surface-sensitive and provides a reliable spectrum for a small amount of material.

Various optical properties, including optical conductivity, can be obtained from measured reflectance spectra of cuprates\cite{wooten,tanner:2019}. Information on the pairing interaction for forming the Cooper pairs can be encoded in the optical conductivity of cuprates via the band renormalization caused by the strong correlations between electrons. The extended Drude model formalism\cite{webb:1986,puchkov:1996,hwang:2004} has been used to decode the information from the measured optical conductivity. The optical self-energy defined by the extended Drude model formalism contains information on the correlations between electrons\cite{hwang:2004}. The optical self-energy can be described in terms of the EBSD function; these two quantities can be related via so-called Allen formulas\cite{allen:1971}, which are integral equations. Numerical methods, such as the least squares fitting method or the maximum entropy method (MEM)\cite{hwang:2006,schachinger:2006,hwang:2007} have been used to obtain the EBSD function from a measured optical self-energy by solving this inversion problem. The least squares fitting method is model-dependent, whereas the MEM is model-independent. However, the amplitude of the EBSD function obtained by using the MEM is not uniquely determined because experimental spectra naturally contain random noise\cite{hwang:2016a}; its amplitude increases as the fitting quality improves. Therefore, the amplitude of the EBSD function obtained by using the MEM may have uncertainty, which makes it difficult to compare the EBSD functions obtained using different experimental techniques such as optical, tunneling, ARPES, and inelastic neutron scattering. A recent study on the analytic continuation problem, which is an inversion problem, using machine learning showed that a machine-learning-based approach provides a more accurate resulting spectrum than the conventional MEM and is more robust against noise in terms of peak positions and amplitude\cite{yoon:2018}.

In this study, to solve the uncertainty problem in the amplitude of the EBSD function obtained by the MEM, we developed a new method to obtain EBSD functions from measured optical spectra using machine learning. We generated 110,000 optical scattering rates with model EBSD functions consisting of a Gaussian peak, a sharp mode, and a broad Millis-Monien-Pines (MMP) mode\cite{millis:1990} using a generalized Allen formula, which was developed by Shulga {\it et al.}\cite{shulga:1991} and could be used for optical spectra at finite temperatures and with a constant density of states. We also included random noise in the optical scattering rates to make them more realistic. We used 100,000 data for the training and the remaining 10,000 data for the evaluations. We found that the developed machine learning approach (MLA) is quite robust against random noise. To further verify our MLA, we applied it to existing measured optical scattering rates at 100 K for one optimally doped ($T_c$ = 96 K) and two overdoped ($T_c$ = 82 K and 60 K) Bi$_2$Sr$_2$CaCu$_2$O$_{8+\delta}$ (Bi-2212) samples. We were able to obtain EBSD functions with reasonable amplitudes from the measured spectra of Bi-2212. We also found that the MLA depends on the training data set (or model EBSD functions); this issue is discussed in the Supplementary Materials\cite{sm:2021a}.

\section{Machine learning approach for inversion problems}
A successful MLA requires a training data set, that is rich enough to account for data manifold, and a machine learning model (e.g., a deep learning network), that is capable of learning meaningful features hidden in the training data set. Finally, a well-trained model should produce a reasonable output for a new data set, that has not been used in the training. Here, we developed a novel MLA to solve the inverse problem of getting the EBSD function $I^2\chi(\omega)$ from a measured optical scattering rate $1/\tau^{op}(\omega)$. Here, $I$ is the coupling constant between an itinerant electron and a force-mediating boson and $\chi(\omega)$ is the boson spectrum. The generalized Allens's formulas can be written in the following equation as
\begin{equation}
  1/\tau^{op}(\omega) = F(I^2\chi(\omega)),
\end{equation}
where $F(\cdot)$ represents a forward (integral) operator. The MLA learns to mimic its inverse operation $F^{-1}: 1/\tau^{op}(\omega) \rightarrow I^2\chi(\omega)$, via the machine. Because the data space is huge in this case due to discretization, it is hard to get a proper training data set that could represent the data manifold of our problem without experts’ domain knowledge. In this regard, the training data sets can be a constraint for applications of the MLA. Therefore, one needs to appropriately design a training data set for his/her input data to be analyzed.

In general, the electron-boson spectral density function [$I^2\chi(\omega)$] of cuprates can be obtained from measured reflectance spectrum by using a well-established process\cite{schachinger:2006,hwang:2015a}. This process consists of a series of steps from measured reflectance, through the optical conductivity, the optical self-energy (or optical scattering rate), to the EBSD function\cite{hwang:2015a}. In this paper, we focus on the last step of the process from the optical scattering rate [$1/\tau^{op}(\omega,T)$] to the EBSD function [$I^2\chi(\omega,T)$] using one of the generalized Allen formulas\cite{shulga:1991} as
\begin{eqnarray}
  \frac{1}{\tau^{op}(\omega, T)} &=& \int^{\infty}_{0} d\Omega\: I^2\chi(\Omega,T) K(\omega, \Omega, T), \\  \nonumber
  K(\omega, \Omega, T) &=& \frac{\pi}{\omega}\Big{[} 2\omega \coth\Big{(}\frac{\Omega}{2T} \Big{)} - (\omega + \Omega) \coth \Big{(} \frac{\omega+\Omega}{2T} \Big{)}  \\  \nonumber
  &+& (\omega - \Omega) \coth \Big{(} \frac{\omega-\Omega}{2T} \Big{)} \Big{]},
\end{eqnarray}
where $T$ is the temperature and $K(\omega, \Omega, T)$ is the kernel, which depends on the material phases\cite{hwang:2018}. The kernel in Eqn. (2) is known as the Shulga's kernel\cite{shulga:1991} and can be applied to optical spectrum at a finite temperature and with a constant density of states. The integral Allen formula needs to be solved through inversion methods to obtain the EBSD function from a measured optical scattering rate. As mentioned previously, the model-dependent least squares fitting method and the model-independent MEM method have been used to numerically solve the integral equation\cite{hwang:2006,schachinger:2006,hwang:2007}. In this study, we developed a new method to obtain the EBSD function from measured optical scattering rate via an MLA.

Our approach is based on the recent developments in machine learning, especially in the area of deep learning \cite{lecun:2015}. With the advent of big data and unprecedented computational power, deep learning has been applied in various disciplines in science and engineering and has achieved impressive successes\cite{hannun:2014, yamins:2014, sadowski:2014}. Deep learning utilizes multiple interconnected layers, which allow for representing rich nonlinear models\cite{SIEGELMANN1995}. With an ample amount of data, a well-trained deep learning network produces much more accurate and faster results than any conventional approaches in a wide range of problems. For a more detailed information on machine learning and its application to physics, we refer the reader to a literature\cite{Carleo:2019}.

\begin{figure}[!htbp]
  \vspace*{-1.0 cm}%
  \centerline{\includegraphics[width=5.5 in]{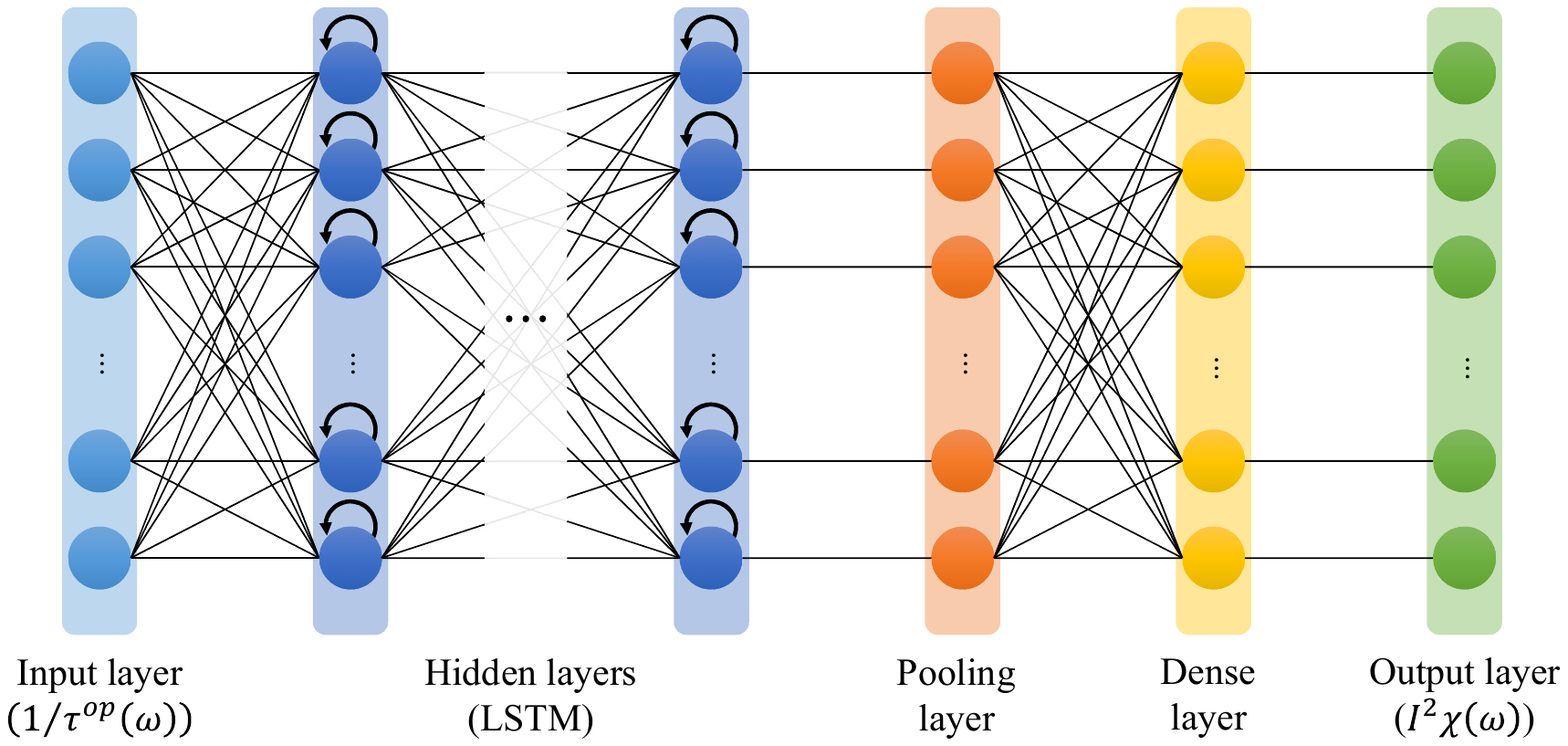}}%
  \vspace*{-1.5 cm}%
\caption{(Color online) A schematic diagram of the neural network structure with a long short-term memory (LSTM) layers for deep learning.}
 \label{fig1}
\end{figure}

For this study, we applied deep learning to solve Eqn. (2). Because our goal is to obtain the EBSD function [$I^2\chi(\omega, T)$] from the optical scattering rate [$1/\tau^{op}(\omega,T)$], this can be cast as an inverse problem. Owing to the ill-posedness of the inverse problem, its solution is well known to be sensitive to noise, as demonstrated in a literature\cite{hwang:2016a}. Some recent studies have solved ill-posed inverse problems using deep learning \cite{Adler:2017, Senouf:2019}; however, they are concerned with images using convolutional neural networks (CNNs). We adopted the long short-term memory (LSTM) layers \cite{hochreiter:1997} into our model because our input [$1/\tau^{op}(\omega,T)$] can be considered as a sequential data and the LSTM network is best suited for such a case. Whereas in a typical deep neural network with convolutional layers the process passes through each layer only once, in the LSTM the process is repeated before it goes to the next layer, as marked with the black circular arrows in the hidden layers of Fig. \ref{fig1}. We used three LSTM hidden layers for this study. In the Supplementary Materials\cite{sm:2021a}, we show that the model with the LSTM layers performs better than that with a CNN.

\section{Results and discussions}

\begin{figure}[!htbp]
  \vspace*{-0.5 cm}%
  \centerline{\includegraphics[width=4.3 in]{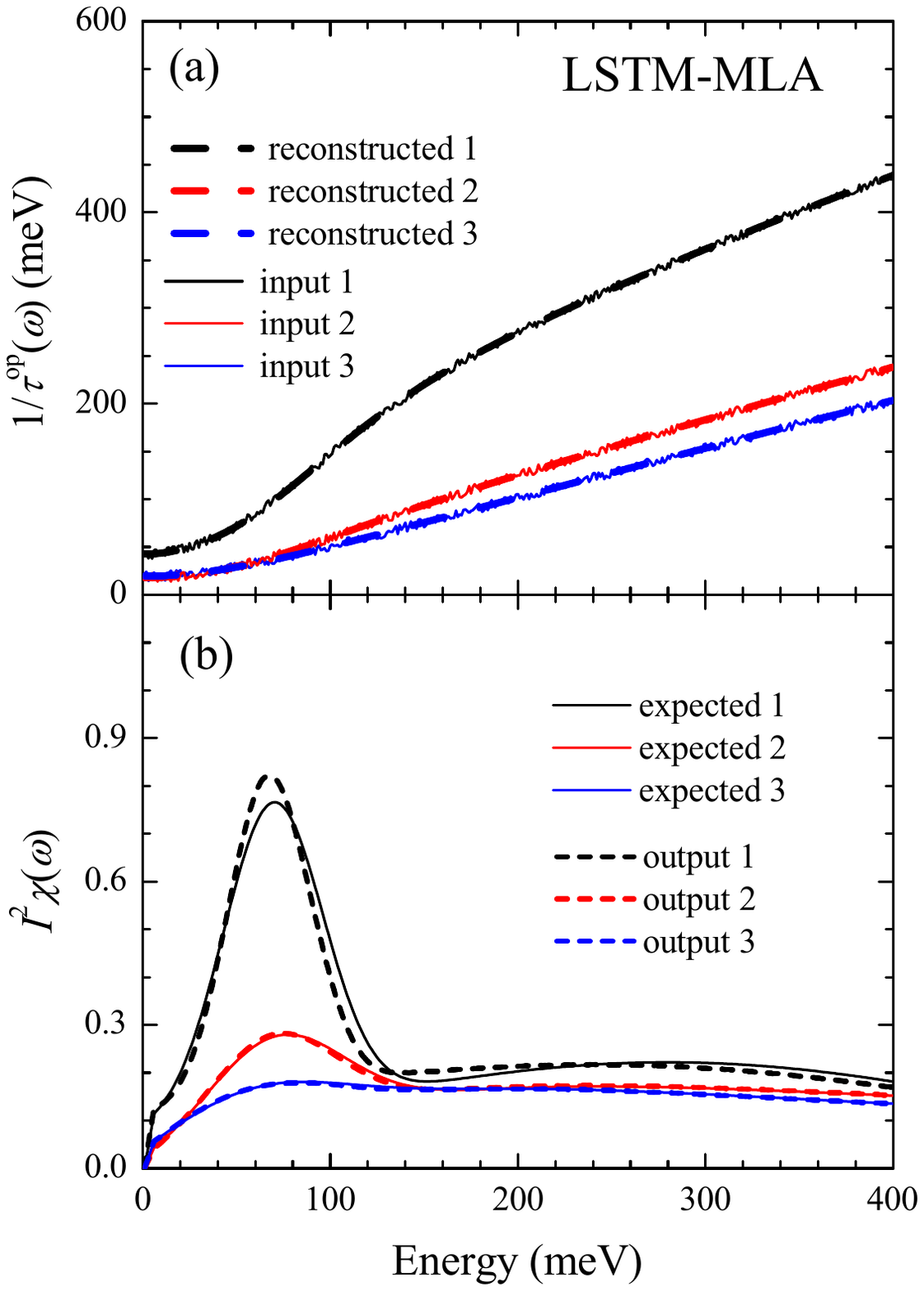}}%
  \vspace*{-0.7 cm}%
\caption{(Color online) Validation of machine learning. (a) Input data added random noise with the amplitude of 2.5 meV (solid lines) and the corresponding ones (dashed lines) reconstructed using the long short-term memory machine learning approach (LSTM-MLA). (b) Comparison of the $I^2\chi(\omega)$ (dashed line) obtained using machine learning approach with the $I^2\chi(\omega)$ (solid lines) in the original (or expected) data set.}
 \label{fig2}
\end{figure}

The training data were generated using Eqn. (2). They consist of three columns: frequency, input data [$1/\tau^{op}(\omega,T)$], and output data [$I^2\chi(\omega,T)$], where $T$ = 100 K. To generate the input data $1/\tau^{op}(\omega,T)$, we used a model output $I^2\chi(\omega,T)$, which consists of the following three terms:
\begin{eqnarray}
I^2\chi(\omega,T) &=& \frac{A_p(T)}{\sqrt{2\pi} [d(T)/2.35]}\exp{\Big{\{} -\frac{[\omega-\omega_p(T)]^2}{2[d(T)/2.35]^2} \Big{\}}} \\ \nonumber &+& \frac{A_s(T) \omega}{\omega^4+[\omega_s(T)]^4} + \frac{A_m(T) \omega}{\omega^2+[\omega_m(T)]^2},
\end{eqnarray}
where the first term is a Gaussian peak located at $\omega_p(T)$ with an amplitude of $A_p(T)$ and a width of $d(T)$ and the second and third terms were previously used for analyzing underdoped cuprates\cite{hwang:2011}. The third term is known as the MMP mode\cite{millis:1990}, which was used for describing antiferromagnetic fluctuations. The first and second terms are sharp components, whereas the third one is a broad component. It is worth noting that the model output $I^2\chi(\omega)$ depends on both temperature\cite{hwang:2016a} and doping\cite{hwang:2018}. We generated $110,000$ data by systematically changing all the parameters within certain ranges. The first $100,000$ data were used for the training and the rest for the validation in each iteration. We used the Adamax optimizer\cite{kingma:2014}, which is a stochastic gradient descent method and optimized on the basis of the square of the exponential value of the slope. We also tried Adam\cite{kingma:2014} and Adadelta\cite{zeiler:2012} but found that Adamax provided the best result. The mean square error (MSE) and the softplus function were used as the loss and the activation functions, respectively, for the training\cite{dugas:2001}. To make the input data [$1/\tau^{op}(\omega)$] more realistic, we added random noises with an amplitude of 2.5 meV. The training took approximately $19$ h on an NVIDIA GeForce RTX 2070 Super graphics processing unit (GPU). The MSE of the validation using the remaining $10,000$ data was 8.6$\times 10^{-5}$, which indicates that the machine was well-trained. It took only $\sim$5 ms to obtain a solution with the trained model. We also performed the training and validation processes using the LSTM method with different ratios of validation ($N_{\mathrm{valid}}$) to training ($N_{\mathrm{train}}$) data sets (i.e., $N_{\mathrm{valid}}/N_{\mathrm{train}}$) and obtained similar results for each case (see Supplementary Information\cite{sm:2021a} for a detailed discussion).

\begin{figure}[!htbp]
  \vspace*{-0.5 cm}%
  \centerline{\includegraphics[width=4.3 in]{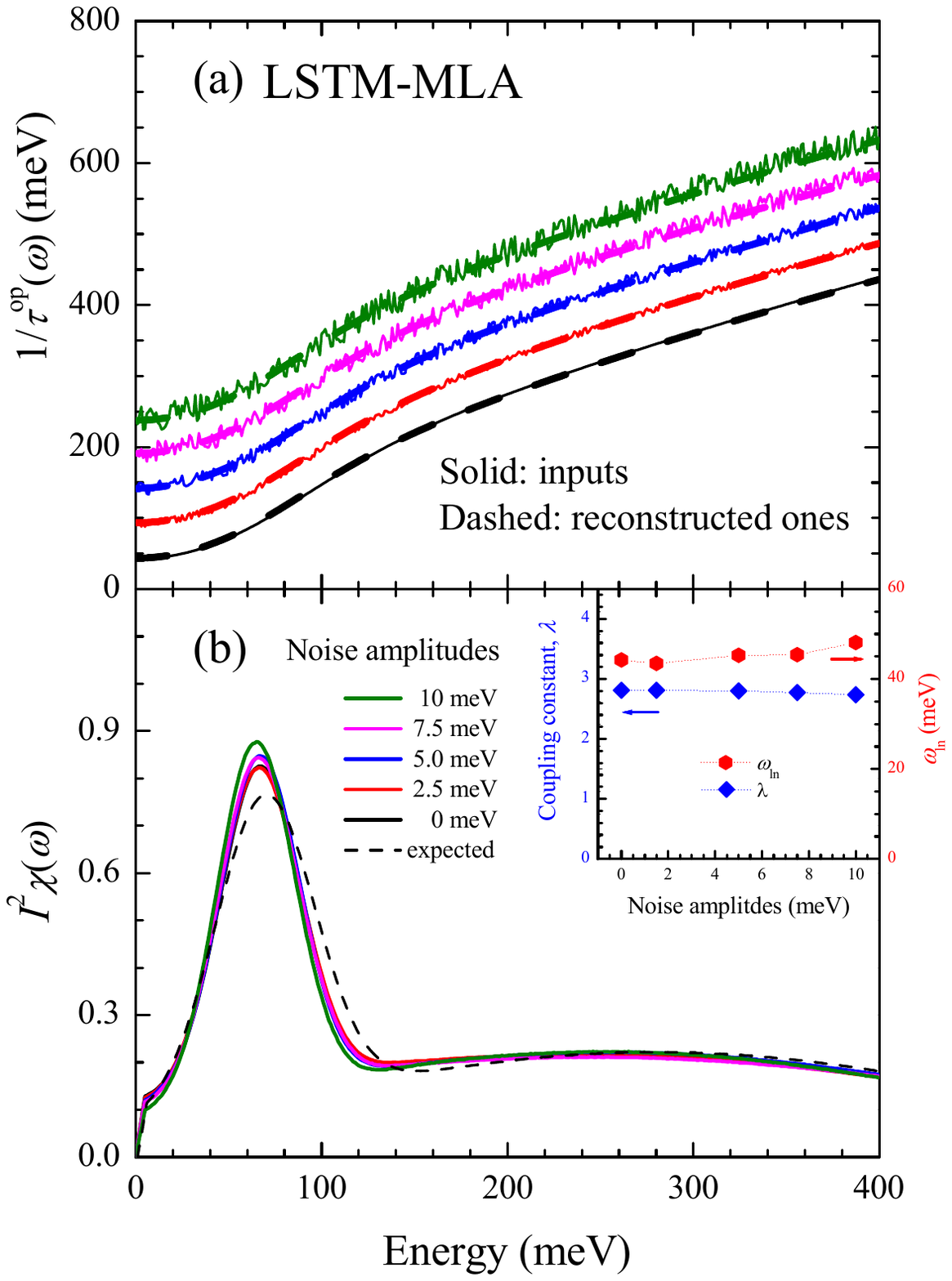}}%
  \vspace*{-0.8 cm}%
\caption{(Color online) Robustness of the LSTM-MLA against random noises. (a) The input $1/\tau^{op}(\omega)$ with various amplitudes of random noise and the corresponding reconstructed input $1/\tau^{op}(\omega)$. (b) The resulting output $I^2\chi(\omega)$ obtained from the input data with different levels of random noise by using the LSTM-MLA. In the inset, noise amplitude dependent coupling constant and logarithmically averaged frequency are shown.}
 \label{fig3}
\end{figure}

In Fig. \ref{fig2}(a), we illustrate the three input data (solid lines) and the corresponding data (dashed lines) reconstructed using the long short-term memory-machine learning approach (LSTM-MLA). They agree well with each other. In Fig. \ref{fig2}(b), we depict the resulting output $I^2\chi(\omega)$ (dashed lines) obtained from the three input data $1/\tau^{op}(\omega)$ using the LSTM-MLA and the original $I^2\chi(\omega)$ (solid lines) used for generating the evaluation data. Overall agreements are quite good. On the basis of this evaluation, we concluded that the training was performed well. We note that in the case of input 1, which contains a sharp increase, the output 1 and the expected 1 exhibit small disagreement near the peak, indicating that the sharp increase in $1/\tau^{op}(\omega)$ may not be sensitive enough to the shape of the sharp peak in the model $I^2\chi(\omega)$. This is not associated with random noise; the output $I^2\chi(\omega)$ for noise amplitudes of 0 and 2.5 meV are the same, as shown in Fig. \ref{fig3}(b).

\begin{figure}[!htbp]
  \vspace*{-0.7 cm}%
  \centerline{\includegraphics[width=4.7 in]{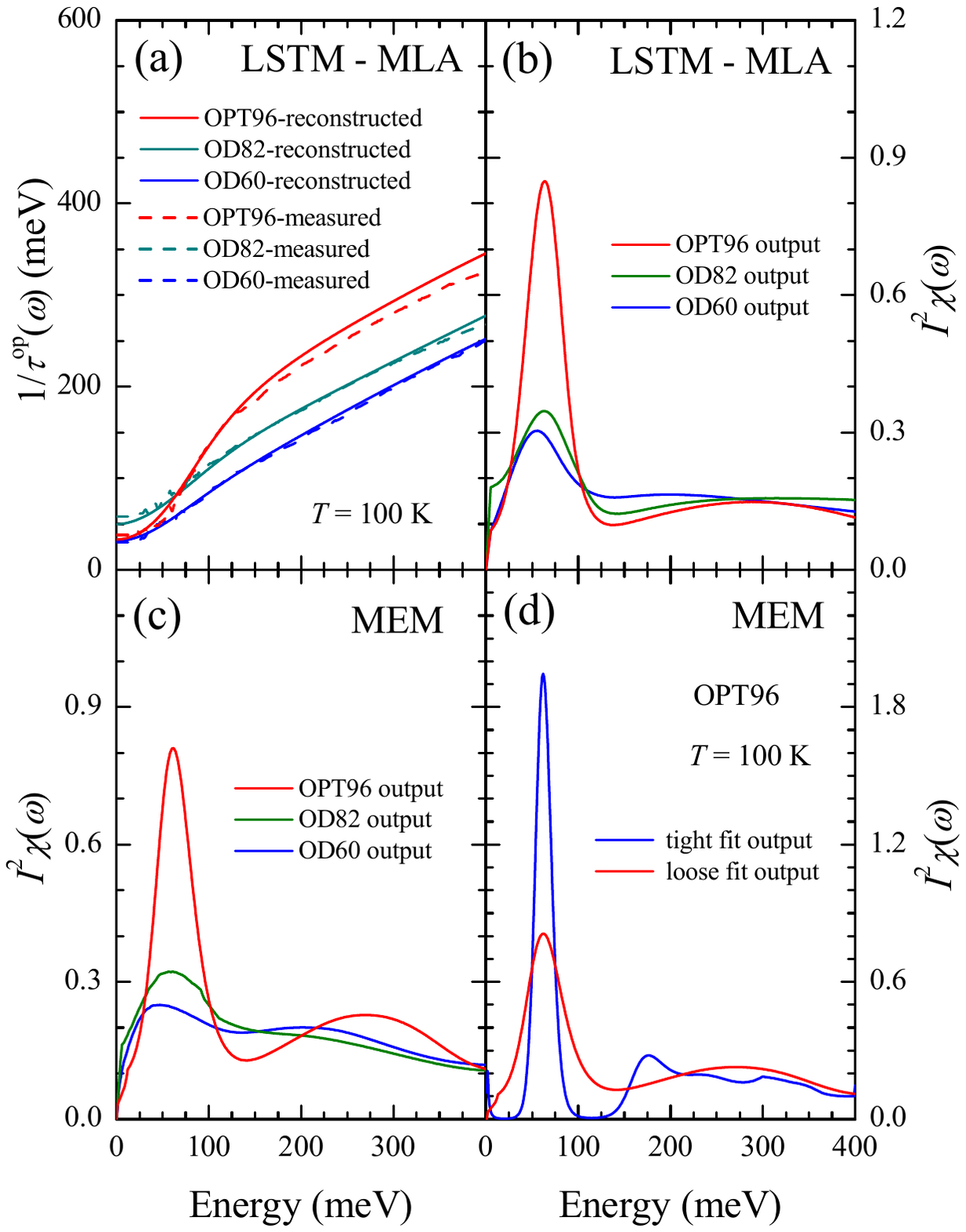}}%
  \vspace*{-0.5 cm}%
\caption{(Color online) Applications of the developed LSTM-MLA to experimentally measured optical spectra. (a) The existing experimental optical scattering rates of Bi-2212 samples (OPT96, OD82, and OD60) and the corresponding input optical scattering rates reconstructed using the LSTM-MLA. (b) The resulting EBSD functions obtained from the measured optical scattering rates using the LSTM-MLA. (c) The EBSD functions obtained from the measured optical scattering rates using the MEM. (d) The EBSD functions obtained from the measured optical scattering rate of OPT96 using the MEM with two different fitting quality levels.}
 \label{fig4}
\end{figure}

Furthermore, we investigated the robustness of the resulting output $I^2\chi(\omega)$ obtained using the LSTM-MLA against random noise. We additionally generated the input $1/\tau^{op}(\omega)$ (solid lines) with various amplitudes (from 0 up to 10 meV) of random noise and the corresponding data (dashed lines) reconstructed using the LSTM-MLA, as shown in Fig. \ref{fig3}(a). The curves are progressively shifted by 50 meV from the curve with no noise for clarity. We obtained the output $I^2\chi(\omega)$ from the generated input $1/\tau^{op}(\omega)$ using the LSTM-MLA. In Fig. \ref{fig3}(b), we depict the output $I^2\chi(\omega)$ obtained from the input data using the LSTM-MLA. We also show the expected (or original) $I^2\chi(\omega)$ (a dashed line). The resulting output data show almost no noise-level dependence; there are few changes in the position, height, and shape of the peak, regardless of the noise levels. But, if we closely investigate the differences, the higher noise-level gives the more discrepancy compared with the expected one. However, the overall shape of the output is quite robust against random noise, which is stark contrast to the results obtained using the MEM\cite{hwang:2016a} (also see Fig. \ref{fig4}(d)). We further analyzed the EBSD functions obtained from inputs with different noise amplitudes using the LSTM-MLA. We calculated the coupling constant ($\lambda$), which is defined as $\lambda \equiv 2\int_{0}^{\omega_c}d\Omega [I^2\chi(\Omega)/\Omega]$, where $\omega_c$ is a cutoff frequency. Here, we used the cutoff frequency of 400 meV. We also calculated the logarithmically averaged frequency, which is defined as $\omega_{ln} \equiv \exp{\{(2/\lambda) \int_{0}^{\omega_c}d\Omega \ln{\Omega} \: [I^2\chi(\Omega)/\Omega] \}}$. These two quantities are important to estimate the superconducting transition temperature\cite{carbotte:1990} and were reported to be robust to fitting quality\cite{hwang:2016a}. The calculated $\lambda$ and $\omega_{ln}$ from the obtained $I^2\chi(\omega)$ using the LSTM-MLA were showed in the inset of Fig. \ref{fig3}(b). Both $\lambda$ and $\omega_{ln}$ showed small noise-amplitude dependencies up to 7.5 meV. For the case of the noise amplitude of 10 meV, we observed some deviation from compared with that of no noise case. From the further analyses above and an earlier study\cite{hwang:2016a}, we could see that the two quantities ($\lambda$ and $\omega_{ln}$) were not very sensitive to the amplitude (or correct shape) of the EBSD function. However, the correct-shaped EBSD function can be crucial to figure out the underlying superconducting pairing mechanism by comparing it with results of other spectroscopic experimental techniques such as ARPES, STM, and INS and for designing theoretical models.

Then, we applied the LSTM-MLA to the existing experimentally measured optical spectra: the optical scattering rates of one optimally doped ($T_c =$ 96 K) and two overdoped ($T_c =$ 82 and 60 K) Bi-2212 samples at $T$ = 100 K. We denote these measured samples as OPT96, OD82, and OD60, respectively\cite{hwang:2007}. Because these samples have negligible pseudogap, we used the generalized Allen formula in Eqn. (2) to analyze these spectra. In Fig. \ref{fig4}(a), we depict the three measured optical scattering rates at 100 K and the corresponding $1/\tau^{op}(\omega)$ spectra reconstructed using the developed LSTM-MLA. Fig. \ref{fig4}(b) shows the resulting output EBSD functions obtained from the three measured optical scattering rates using the LSTM-MLA. In Fig. \ref{fig4}(c), we also show the EBSD functions obtained from the optical scattering rates of the same samples using the MEM, which were previously reported in the literature\cite{hwang:2007}. The corresponding EBSD functions in Fig. \ref{fig4}(b) and (c) look similar to each other. However, if the fitting quality is improved in the MEM, the shape of the EBSD function, including peak height, is significantly changed, as shown in Fig. \ref{fig4}(d), where we illustrate the EBSD functions obtained from the optical scattering rate of OPT96 using the MEM with two different fitting quality-levels\cite{hwang:2016a}. The two resulting EBSD functions are significantly different, except for the position of the sharp peak located near 50 meV. Therefore, the shape of the EBSD function obtained using the MEM is not uniquely determined because of random noise, which exists naturally in experimentally measured optical spectra\cite{hwang:2016a}. However, as noted previously, the LSTM-MLA is highly robust against random noise and facilitates obtaining a correct-shaped EBSD function from the measured optical scattering rate, which naturally contains random noise.

Thus far, we have developed an MLA to obtain EBSD functions from measured optical scattering rates for the case of the normal state at $T =$ 100 K. We can easily extend the LSTM-MLA for other temperatures in the normal state. In the cases of the normal state with a pseudogap and the superconducting state, it will take much more time to generate the training data with the same data points because, for these two cases, the generalized Allen formulas consist of double integrations\cite{sharapov:2005,schachinger:2006,hwang:2008b}. However, for any case, as long as sufficient training data are generated, an MLA can be simply developed using a deep learning neural network.

We found an important issue on the MLA for solving the inverse problem. Approximate information on the shape of the output function needs to be known a priori in order to generate appropriate training data. If a training data set with highly different shapes from those of the measured input spectra is generated, the corresponding input data reconstructed using the MLA deviate significantly from the measured input data. Consequently, the output EBSD function obtained using the MLA will not be a reliable EBSD function for the measured input data. In fact, the shape of the resulting output EBSD function will be close to that of the model EBSD function, which is used for generating the training data. Therefore, the MLA is, in this regard, model-dependent. To explicitly demonstrate this model-dependent issue of the MLA, we generated a training data set with the model output $I^2\chi(\omega)$ consisting of only the last two terms in Eqn. (3) and developed an LSTM-MLA. Then, we applied the developed LSTM-MLA to the measured optical scattering rates of Bi-2212 (OPT 96, OD82, and OD60). The results are presented in the Supplementary Materials\cite{ms:2021a}; the LSTM-MLA is evidently model-dependent. Therefore, to generate our training data in this study, we utilized the shapes of the EBSD functions obtained from the measured optical scattering rates of Bi-2212 using the MEM in a previously reported paper\cite{hwang:2007}. At this stage, one may ask what are advantages of the MLA compared with the least squares fitting method (LSFM) because both approaches are model-dependent. In fact, the LSFM has much less number of fitting parameters. However, the MLA can be potentially extended and eventually a model-independent MLA will be developed in the future.

\section{Conclusion}
We developed a new method to obtain the EBSD function from the optical scattering rates using the MLA. We generated 100,000 training data from the model EBSD functions using the generalized Allen formula. For the training, we used a deep learning neural network with LSTM hidden layers. We compared the results with those obtained using a deep learning neural network with a CNN. We found that the LSTM-MLA took a longer time to train but provided more accurate and stable results compared with the CNN-MLA. We found that the MLA was quite robust against random noise. We applied the developed MLA to the existing experimentally measured optical data of Bi-2212 and obtained EBSD functions with reasonable shapes, including their amplitudes. An earlier study showed that the MLA can be used to solve an inverse problem and is robust against random noise\cite{yoon:2018}. There was an attempt to expose hidden self-energies of cuprates from measured ARPES spectra using an MLA\cite{yamaji:2020}. However, to the best of our knowledge, our study is the first application of the MLA to obtain the EBSD function from measured infrared/optical spectra. From the applications, we found that the MLA is model-dependent, as the results depend on the shapes of the model EBSD functions used for generating the training data. We expect that the MLAs are useful and rapid methods for solving other inversion problems, which may contain random noise. We believe that model-independent MLAs will be developed in the near future for their wide and useful applications to analyses of various regression problems including inverse problems. One possible approach would be using invertible neural networks (INNs)\cite{Ardizzone2019-oz}, where a forward model such as Eqn. (1) is adopted during training. INNs tries to learn the hidden representation of the forward model, which is usually lost in the conventional MLA.

%
\acknowledgments This paper was supported by the National Research Foundation of Korea (NRFK Grant No. 2017R1A2B4007387, 2019R1A6A1007307912, and 2021R1A2C101109811).

\bibliographystyle{apsrev4-1}
\bibliography{bib}

\end{document}